\documentclass[10 pt, letter]{article}
\usepackage{graphicx}
\usepackage{caption}
\usepackage{amsmath}
\usepackage{amsthm}
\usepackage{enumerate}
\usepackage{subfig}
\usepackage{float}
\usepackage{amsfonts}
\usepackage{amssymb}
\usepackage{anysize}
\usepackage{helvet}
\usepackage{setspace}
\usepackage{multirow}
\usepackage{tabularx}
\usepackage{tabulary}
\usepackage{longtable}
\usepackage{float}
\usepackage{mathtools}
\newcommand{\cor}[1]{\left[ #1 \right]}             
\newcommand{\lla}[1]{\left\{ #1 \right\}}           
\newcommand{\tn}[1]{{\bf #1}}                       
\newcommand{\ti}[1]{{\it #1}}                       

\begin{document}

\title{\bf Study of decoherence of entangled states made up of two basic states in a linear chain of three qubits}
\author{G.V. L\'opez\footnote{gulopez@udgserv.cencar.udg.mx}~~Êand~ G.C. Montes\\
\small \emph{Departamento de F\'{i}sica, Universidad de Guadalajara,}\\
\small \emph{Blvd. Marcelino Garc\'{i}a Barragan y Calzada Ol\'{i}mpica, 44200 Guadalajara, Jalisco, Mexico}}
\date{\ \\ \today}
\maketitle

\begin{abstract}
Using Lindblad approach to study decoherence of quantum systems, we study the decoherence and decay of entangled states, formed by  two basic states of a chain of thee qubits. We look on these states for a possible regular dependence on  their decay as a function of their energy separation between the basic states under  different type of environments. We found not regular or significant dependence on this energy separation for the type of environment considered . 
\end{abstract}

\centerline{PACS:  03.65.Yz, 03.67.Bg, 03.67.Mn, }

\section{Introduction } 
In the real world (quantum or classical) the interaction of the system with the environment is unavoidable. In principle, one could study the unitary evolution of the whole system, quantum plus environment plus quantum-environment interaction, but this represent a many bodies problem which is unsolvable within any picture of the quantum mechanics. The most used approach to study this phenomenom is to use the matrix density approach for the whole system and to make the trace over the environment variables \cite{CaLeg,HuPazZh,Venu,Bre_OQS,Haro}.
The resulting density matrix is called "reduced density matrix", and its associated non-unitary evolution equation is called "master equation." This equation is  phenomenological where a dissipative and diffusion parameters are defined, and they are responsible of the decay behavior of the non diagonal elements of the reduced density matrix.  This phenomenon is called "decoherence" because is related with the disappearance of the interference terms of the product of the quantum wave function \cite{ZuP,Zu2} 
and many think that this decoherence effect is closely related with the appearance of the classical world \cite{Zeh,Zu1,Zu2}. 
In most of the approaches, the positiveness and trace equal to one are kept as principal condition for the reduced density matrix. The best known mathematical approach which kept these conditions was given by Lindblad \cite{G_Lim},
which gave an abstract  general non unitary evolution equation  for the reduced density matrix. The master equation is different when dealing with continuos systems (quantum Browning motion, for example)~\cite{CaLeg, HuPazZh} or discrete quantum systems (spin system)~\cite{All}. One of the used approaches  for quantum discrete system is described in ~\cite{Suman}, and we will use this approach for our study of decoherence of entangled states built up with two states of three qubits in a quantum computer model of a linear chain of three paramagnetic atoms with nuclear spin one half~\cite{Berman_02}. 
 In this work, we are interested in determine the decoherence of several entangled states formed by two states of three qubits, and we will use
 the above mentioned Markovian-Lindblad master type of equation \cite{Bre_OQS,Suman}. 
 On the other hand, even this model for solid state quantum computer has not been built, it has been very useful for theoretical studies about implementation of quantum gates and quantum algorithms~\cite{Berman_02,Lop_03,Lop_08,Lop_12} which can be extrapolated to other solid state quantum computers. The main idea is to explore the possible sensitivity of the decay of an entangled state with respect the difference energy of its two states involved, we establish the four cases to be considered with the quantum-environment system: independent environment interaction, pure dephasing interaction, correlated dissipation interaction, and dephasing correlated interaction. The analytical dynamical systems  of the reduced density matrix elements   are obtained for these cases, and the results of the analytical and numerical simulations are presented. 
\section{Hamiltonian of the chain of nuclear spins}
Following Lloyd's idea \cite{Lloyd_93}, consider a linear chain of nuclear spin one half, separated by some distance and inside a magnetic in a direction $z$, ${\bf B}(z)=(0,0,B_0(z))$,  and making and angle $\theta$ with respect  this linear chain. Choosing this angle such that $\cos\theta=1/\sqrt{3}$, the dipole-dipole interaction is canceled, the Larmore's frequency for each spin is different, $\omega_k=\gamma B_0(z_k)$ with 
$\gamma$  the gyromagnetic ratio. The magnetic moment of the nucleus $\vec\mu_k$ is related with its spin through the relation $\vec\mu_k=\gamma {\bf S}_k$, and the interaction energy between the magnetic field and magnetic moments is $H_{int}=-\sum_k\vec\mu_k\cdot{\bf B}(z_k)=-\sum_k\omega_kS_k^z$. If in addition, one has first and second neighbor Ising interaction, the Hamiltonian of the system is just \cite{Berman_02}
\begin{equation}
H_s=-\sum_{k=1}^N\omega_kS_k^z-\frac{2J}{\hbar}\sum_{k=1}^{N-1}S_k^zS_{k+1}^z-\frac{2J'}{\hbar}\sum_{k=1}^{N-2}S_k^zS_{k+2}^z,
\end{equation} 
where $N$ is the number of nuclear spins in the chain (or qubits), $J$ and $J'$ are the coupling constant of the nucleus at first and second neighbor. Using the basis of the register of N-qubits, $\{|\xi_{_N},\dots,\xi_1\rangle\}$ with $\xi_k=0,1$, one has that $S_k^z|\xi_k\rangle=(-1)^{\xi_k}\hbar|\xi_k\rangle/2$. Therefore, the Hamiltonian is diagonal on this basis, and its eigenvalues are
\begin{equation}
E_{\xi}=-\frac{\hbar}{2}\sum_{k=1}^N(-1)^{\xi_k}\omega_k-\frac{J\hbar}{2}\sum_{k=1}^{N-1}(-1)^{\xi_k+\xi_{k+1}}-\frac{J'\hbar}{2}\sum_{k=1}^{N-2}(-1)^{\xi_k+\xi_{k+2}}.
\end{equation}
\section{Interaction with the environment}
Consider now that the environment is characterized by a Hamiltonian $H_e$ and its interacting with the quantum system with Hamiltonian $H_s$. Thus,  the total Hamiltonian would be $H=H_s+H_e+H_{se}$, where $H_{se}$ is the part of the Hamiltonian which takes into account the interaction system-environment, and the equation one would need to solve, in terms of the density matrix, is ~\cite{Fano,von}
\begin{equation}
i\hbar\frac{\partial \rho_t}{\partial t}= [H,\rho_t],
\end{equation} 
where $\rho_t=\rho_t(s,e)$is the density matrix which depends on the system and environment coordinates. The evolution of the system is unitary, but it is not possible to solve this equation. Therefore, under some approximations and tracing over the environment coordinates ~\cite{Davies,Haro}-, it is possible to arrive to a Lindblad type of equation ~\cite{Bre_OQS,Alicki} for the reduced density matrix $\rho(s)=tr_e(\rho_t)$,
\begin{equation}
i\hbar\frac{\partial \rho}{\partial t} = [H_s,\rho] + \sum_{i=1}^{I}\biggl\{V_i\rho V_{i}^{\dagger} - \frac{1}{2}V_{i}^{\dagger}V_i\rho - \frac{1}{2} \rho V_{i}V_{i}^{\dagger}\biggr\}
\end{equation}
where  $V_i$ are called Kraus' operators. This equation is not unitary and Markovian (without memory of the dynamical process). This equation can be written in the interaction picture, through the transformation
 $    \tilde{\rho} = U\rho U^\dagger$ with $ U = e^{iH_st/\hbar}$,  
 as
\begin{equation}\label{LinU}
  i\hbar \frac{\partial\tilde \rho}{\partial t} = \tilde{\mathcal{L}}(\tilde\rho),
\end{equation}
where $\tilde{\mathcal{L}}(\tilde\rho)$ is the Lindblad  operator
\begin{equation}
\tilde{\mathcal{L}}(\tilde\rho)= \sum_{i=1}^{I} \biggl\{\tilde V_i\tilde\rho \tilde V_{i}^{\dagger} - \frac{1}{2}\tilde V_{i}^{\dagger}\tilde V_i\tilde\rho - \frac{1}{2}\tilde \rho \tilde V_{i}\tilde V_{i}^{\dagger}\biggr\}
\end{equation}
with $\tilde V=UVU^{\dagger}$. The explicit form of Lindblad operator is determined by the type of environment to consider ~\cite{Suman} at zero temperature. In this work we consider dissipation effects and defacing. So, the operators can be    $V_i=S_i^-$ (for dissipation),  $V_i=S_i^z$ (for defacing), and $\gamma_i/i\hbar$ is the coupling constant with the environment.  In this way, one considers the following cases: \\Ê
{\bf (a) Independent :} 
In this case, each qubit of the chain acts independently with the environment, and one has local decoherence of the system. The Lindblad operator is
\begin{equation}
\tilde{\mathcal{L}}(\tilde\rho)  =  \frac{1}{2i\hbar}\sum_{k}^{N}\gamma_{k}\bigl(2\tilde{S}_{k}^{-}\tilde\rho\tilde{S}_{k}^{+} - S_k^+S_k^-\tilde\rho - \tilde\rho \tilde S_k^+\tilde S_k^-\bigr)
\end{equation}
where $\tilde S_k^{+}$ and $\tilde S_k^{-}$ are the ascend and descend operators such that
\begin{equation}
   \tilde{S}_k^{\pm} = U S_k^{\pm}U^{\dagger} =S_k^{\pm}e^{\pm i\hat{\Omega}_k t},
\end{equation}
where $\hat{\Omega}_k$ has been defined as
\begin{equation}
{\hat{\Omega}_k = w_k + \frac{J}{\hbar}(S_{k+1}^z + S_{k-1}^z) + \frac{J'}{\hbar}(S_{k+2}^z + S_{k-2}^z).} 
\end{equation}
{\bf (b) Correlated independent :}
Each qubit interact with the environment but its effect is felt by the other qubits, that  is, the type of interaction is nonlocal with a collective interction between qubits and environment. The Lindblad operator is
\begin{equation}
   \tilde{\mathcal{L}}(\rho)  =  \frac{1}{i\hbar}\sum_{j,k}^{N}\frac{\gamma_{jk}}{2}(2\tilde{S}_{k}^{-}\rho\tilde{S}_{j}^{+} - \tilde{S}_j^+\tilde{S}_{k}^{-}\rho - \rho\tilde{S}_{j}^{+}\tilde{S}_{k}^{-}).
\end{equation}
where  $\gamma_{jk}$ are the coupling constant between qubits and environment,  with $\gamma_{jk} = \gamma_{kj}$  and  $\gamma_{ii}=\gamma_i$ .
\\Ê\\
{\bf (c) Dephasing :}
There is not interchange of energy between qubits and environment, only decoherence is presented where the non diagonal elements of the reduced density matrix go to zero. The Lindblad operator is  
\begin{equation}
   \tilde{\mathcal{L}}(\rho)  =  \frac{1}{i\hbar}\sum_{k}^{N}\Gamma_{k}(2S_{k}^{z}\rho S_{k}^{z} - S_k^zS_k^z\rho - \rho S_k^zS_k^z)
\end{equation}
where $\Gamma_k$ is the parameter of the $kth$- qubit which take into account the dephasing of the qubit with the environment (the tilde operators do not appear due to commutation of this operators with the evolution operator  $U$). \\Ê\\
{\bf (d) Correlated dephasing :}
Here, one takes into account the collective effect of the environment to the qubits. Lindblad's operator is of the form
\begin{equation}
   \tilde{\mathcal{L}}(\rho)  =  \frac{1}{i\hbar}\sum_{j,k}^{N}\Gamma_{jk}(2S_{k}^{z}\rho S_{j}^{z} - S_j^zS_k^z\rho - \rho S_j^zS_k^z),
\end{equation}
where $\Gamma_{jk} = \Gamma_{kj}$ is the parameter with take into account the correlation  ( $\Gamma_{ii} = \Gamma_i$). \\
The analytical solutions for these four cases are given in the appendix.\\Ê\\
\section{\bf Entanglement and GME-concurrence}
Our 3-qubits Hilbert space $H$ is generated by the basis $\{|\xi_1\xi_2\xi_3\rangle\}_{\xi_j=0,1}$. Labeling the qubit of the 3-qubits chain as $ABC$, we understand an entanglement of the form $AB$ when the qubits $\{1,2\}$ are entangled, and we understand an entanglement of the form $ABC$ when the 3-qubits $\{1,2,3\}$ are entangled. The entangled state under our consideration are listen on the table 1.  We chose these state since they are mostly used  on experiments of quantum computation or quantum information.\\Ê\\
\begin{table}[H]
   \centering
   \caption{\small{Entangled states (ordered according their energy separation) }}
   \begin{tabular}{p{5cm}ll}
   Entangled form  &  Initial entangled state                               & $\Delta E/ (\hbar 2\pi MHz)$ \\\hline
                   & & \\
                   &  $|\Psi_{18}\rangle = (|1\rangle+|8\rangle)/\sqrt{2};$ & $\Delta  E_{18} = E_8 - E_1 = 700$\\
   \hspace{1.5cm} ABC&$|\Psi_{27}\rangle = (|2\rangle+|7\rangle)/\sqrt{2};$ & $\Delta  E_{27} = E_7 - E_2 = 500$\\
                   &  $|\Psi_{36}\rangle = (|3\rangle+|6\rangle)/\sqrt{2};$ & $\Delta  E_{36} = E_6 - E_3 = 300$\\
                   &  $|\Psi_{45}\rangle = (|4\rangle+|5\rangle)/\sqrt{2};$ & $\Delta  E_{45} = E_5 - E_4 = 100$\\
                   & & \\
                   &  $|\alpha_{17}\rangle = (|1\rangle+|7\rangle)/\sqrt{2};$ & $\Delta  E_{17}=E_7 - E_1=605.2$\\
   \hspace{1.5cm} AB &$|\alpha_{28}\rangle = (|2\rangle+|8\rangle)/\sqrt{2};$ & $\Delta  E_{28}=E_8 - E_2=594.8$\\
                   &  $|\alpha_{46}\rangle = (|4\rangle+|6\rangle)/\sqrt{2};$ & $\Delta  E_{46}=E_6 - E_4=209.8$\\
                   &  $|\alpha_{35}\rangle = (|3\rangle+|5\rangle)/\sqrt{2};$ & $\Delta  E_{35}=E_5 - E_3=195.2$\\
                   & & \\
                   &  $|\beta_{14}\rangle = (|1\rangle+|4\rangle)/\sqrt{2};$ & $\Delta  E_{14}=E_4 - E_1= 305.2$\\
   \hspace{1.5cm} BC &$|\beta_{58}\rangle = (|5\rangle+|8\rangle)/\sqrt{2};$ & $\Delta  E_{58}=E_8 - E_5= 294.8$\\
                   &  $|\beta_{23}\rangle = (|2\rangle+|3\rangle)/\sqrt{2};$ & $\Delta  E_{23}=E_3 - E_2= 104.8$\\
                   &  $|\beta_{67}\rangle = (|6\rangle+|7\rangle)/\sqrt{2};$ & $\Delta  E_{67}=E_7 - E_6= 95.2$\\
                   & & \\
                   &  $|\xi_{16}\rangle = (|1\rangle+|6\rangle)/\sqrt{2};$ & $\Delta  E_{16}=E_6 - E_1= 510$\\
   \hspace{1.5cm} AC &$|\xi_{38}\rangle = (|3\rangle+|8\rangle)/\sqrt{2};$ & $\Delta  E_{38}=E_8 - E_3= 490$\\
                   &  $|\xi_{25}\rangle = (|2\rangle+|5\rangle)/\sqrt{2};$ & $\Delta  E_{25}=E_5 - E_2= 300$\\
                   &  $|\xi_{47}\rangle = (|4\rangle+|7\rangle)/\sqrt{2};$ & $\Delta  E_{47}=E_7 - E_4= 300$\\
                   & & \\\hline 
   \end{tabular}
\end{table}
\newpage
In order to quantify the entanglement of a state formed by  three qubits basis-states, we will use the criteria given on ~\cite{Huber,Zhi,Seev} where the lower bound of the concurrence is
\begin{equation}
C_{GME}(\Phi) \ge 2\left(\sqrt{\langle{\Phi}|\rho^{\otimes 2}\Pi_{\{1,...,N\}}|{\Phi}\rangle } - \sum_\beta \sqrt{\langle{\Phi}|\Pi_\beta\rho^{\otimes 2}\Pi_\beta|{\Phi}\rangle}\right)
\end{equation}
where  $|\Phi\rangle$ is a separable state of the two copies of the Hilbert space, $\mathcal{H}\otimes \mathcal{H}$.  $\Pi_{\{\alpha\}}$ is the permutation operator acting on the double copies of the Hilbert space, $\mathcal{H}\otimes \mathcal{H}$, interchanging elements of one space into the other, for example $\Pi_{\{1\}}(|\phi_1\phi_2\rangle\otimes|\psi_1\psi_2\rangle) = |\psi_1\phi_2\rangle\otimes|\phi_1\psi_2\rangle$. if  $\zeta_1=\{l|m,n\}$, $\Pi_{\{l|,m,n\}}$ acting on $|\Psi_{rs}\rangle$ means that the qubit label by "$l$" is fixed and the qubits label $"m"$ and $"n"$ are interchanged ($ \Pi_{\{1|2,3\}}|18\rangle  = |45\rangle$, 
$\Pi_{\{1,2,3\}}|18\rangle  = |81\rangle$ ). Denoting by $\rho_{mn} = \langle m|\Psi_{18}\rangle\langle\Psi_{18}|n\rangle$, one has the fallowing GME-concurrence associated to the given entangled state:
\begin{table}[H]
   \centering
   \caption{\small{Entanglement form: $ABC$.}}
   \begin{tabular}{lll}
      State     & $|\Phi\rangle$ &   GME-concurrence \\\hline
      $|\Psi_{18}\rangle$  & $|18\rangle$   &  $2|\rho_{18}| - 2\sqrt{\rho_{44}\rho_{55}} - 2\sqrt{\rho_{33}\rho_{66}} - 2\sqrt{\rho_{22}\rho_{77}}$ \\
      $|\Psi_{27}\rangle$  & $|27\rangle$   &  $2|\rho_{27}| - 2\sqrt{\rho_{11}\rho_{88}} - 2\sqrt{\rho_{33}\rho_{66}} - 2\sqrt{\rho_{44}\rho_{55}}$ \\
      $|\Psi_{36}\rangle$  & $|36\rangle$   &  $2|\rho_{36}| - 2\sqrt{\rho_{11}\rho_{88}} - 2\sqrt{\rho_{22}\rho_{77}} - 2\sqrt{\rho_{44}\rho_{55}}$ \\
      $|\Psi_{45}\rangle$  & $|45\rangle$   &  $2|\rho_{45}| - 2\sqrt{\rho_{11}\rho_{88}} - 2\sqrt{\rho_{22}\rho_{77}} - 2\sqrt{\rho_{33}\rho_{66}}$ \\\hline
   \end{tabular}
\end{table}
\noindent
For the case when entangled state is of the form AB, BC or AC, one makes the trace on the missing letter qubit, and it follows that
( $|\Phi\rangle = | ijkl\rangle$ )
\begin{equation}
\frac{C_{GME}}{2}=|\langle il|\rho_{AB}|kj\rangle| - \sqrt{\langle ij|\rho_{AB}|ij\rangle\langle kl\rangle|\rho_{AB}|kl\rangle},
\end{equation}

\begin{table}[h!]
   \centering
   \caption{\small{Entanglemet form: $AB$.}}
   \begin{tabular}{lll}
      State   & $|\Phi\rangle$ &  GME-concurrence \\\hline
      $|\alpha_{17}\rangle$ & $|0101\rangle$ &  $2|\rho_{17}+\rho_{28}| - 2\sqrt{(\rho_{33}+\rho_{44})(\rho_{55}+\rho_{66})}$ \\
      $|\alpha_{28}\rangle$ &    $\cdots$  &  \hspace{3.5cm}$\cdots$ \\
      $|\alpha_{46}\rangle$ & $|0011\rangle$ &  $2|\rho_{35}+\rho_{46}| - 2\sqrt{(\rho_{11}+\rho_{22})(\rho_{77}+\rho_{88})}$ \\
      $|\alpha_{35}\rangle$ &   $\cdots$   &  \hspace{3.5cm}$\cdots$  \\\hline
   \end{tabular}
\end{table}
\begin{table}[h!]
   \centering
   \caption{\small{Entanglement form: $BC$.}}
   \begin{tabular}{lll}
      State   & $|\Phi\rangle$ &  GME-concurrence \\\hline
      $|\beta_{14}\rangle$ & $|0101\rangle$ &  $2|\rho_{14}+\rho_{58}| - 2\sqrt{(\rho_{22}+\rho_{66})(\rho_{33}+\rho_{77})}$ \\
      $|\beta_{58}\rangle$ &    $\cdots$  &  \hspace{3.5cm}$\cdots$ \\
      $|\beta_{23}\rangle$ & $|0011\rangle$ &  $2|\rho_{23}+\rho_{67}| - 2\sqrt{(\rho_{11}+\rho_{55})(\rho_{44}+\rho_{88})}$ \\
      $|\beta_{67}\rangle$ &   $\cdots$   &  \hspace{3.5cm}$\cdots$  \\\hline
   \end{tabular}
\end{table}
\begin{table}[h!]
   \centering
   \caption{\small{Entanglement form: $AC$.}}
   \begin{tabular}{lll}
      State & $|\Phi\rangle$ &  GME-concurrence \\\hline
      $|\xi_{16}\rangle$ & $|0101\rangle$ &  $2|\rho_{16}+\rho_{38}| - 2\sqrt{(\rho_{22}+\rho_{44})(\rho_{55}+\rho_{77})}$ \\
      $|\xi_{38}\rangle$ &    $\cdots$  &  \hspace{3.5cm}$\cdots$ \\
      $|\xi_{25}\rangle$ & $|0011\rangle$ &  $2|\rho_{23}+\rho_{47}| - 2\sqrt{(\rho_{11}+\rho_{33})(\rho_{66}+\rho_{88})}$ \\
      $|\xi_{47}\rangle$ &   $\cdots$   &  \hspace{3.5cm}$\cdots$  \\\hline
   \end{tabular}
\end{table}
\section{\bf Results}
In our case, we have three qubits space $\{|\xi_2\xi_2\xi_1\rangle\}_{\xi_i=0,1}\}$, and our parameter in units $2\pi MHz$ are
\begin{eqnarray*}
\omega_1    & = & 400; \quad \omega_2     = 200;  \quad \omega_3    = 100    \quad J          = 10;   \quad J'          = 0.4      \\
\gamma_1    & = & 0.05;\quad \gamma_2     = 0.05;  \quad \gamma_3    = 0.05   \quad \Gamma_1  = 0.05; \quad \Gamma_2  = 0.05;  \quad \Gamma_3   = 0.05 \\
\gamma_{12} & =& 0.05;\quad \gamma_{23}  = 0.025 \quad \gamma_{13}  = 0.0125 \quad \Gamma_{12}  = 0.05; \quad \Gamma_{23}  = 0.025; \quad \Gamma_{13}  = 0.0125 \end{eqnarray*}
the time is normalized by the same factor of $2\pi MHz$.
\noindent
To determine the departure of the pure state entangled state, we use the purity parameter, $P=Tr(\rho^2)$ ~\cite{NiCh}. Figure 1a shows the behavior of this parameter for the entangled state $|\Psi_{18}\rangle$ as a function of time, where one can see that correlations does not affect much the independent model of  the environment, which can be seen only for much bigger dissipation parameters, Figure 1b. Dephasing models  finish with the mix state on the system at the end, instead of a pure state of the independent model.  As seen on Figure 2, independent model ends with a pure state in the system due to the system ends on the ground state after sharing energy with the environment. 

\newpage

\begin{figure}[h!]
   \centering
   \subfloat[]{
        \includegraphics[width=8cm,height=8cm]{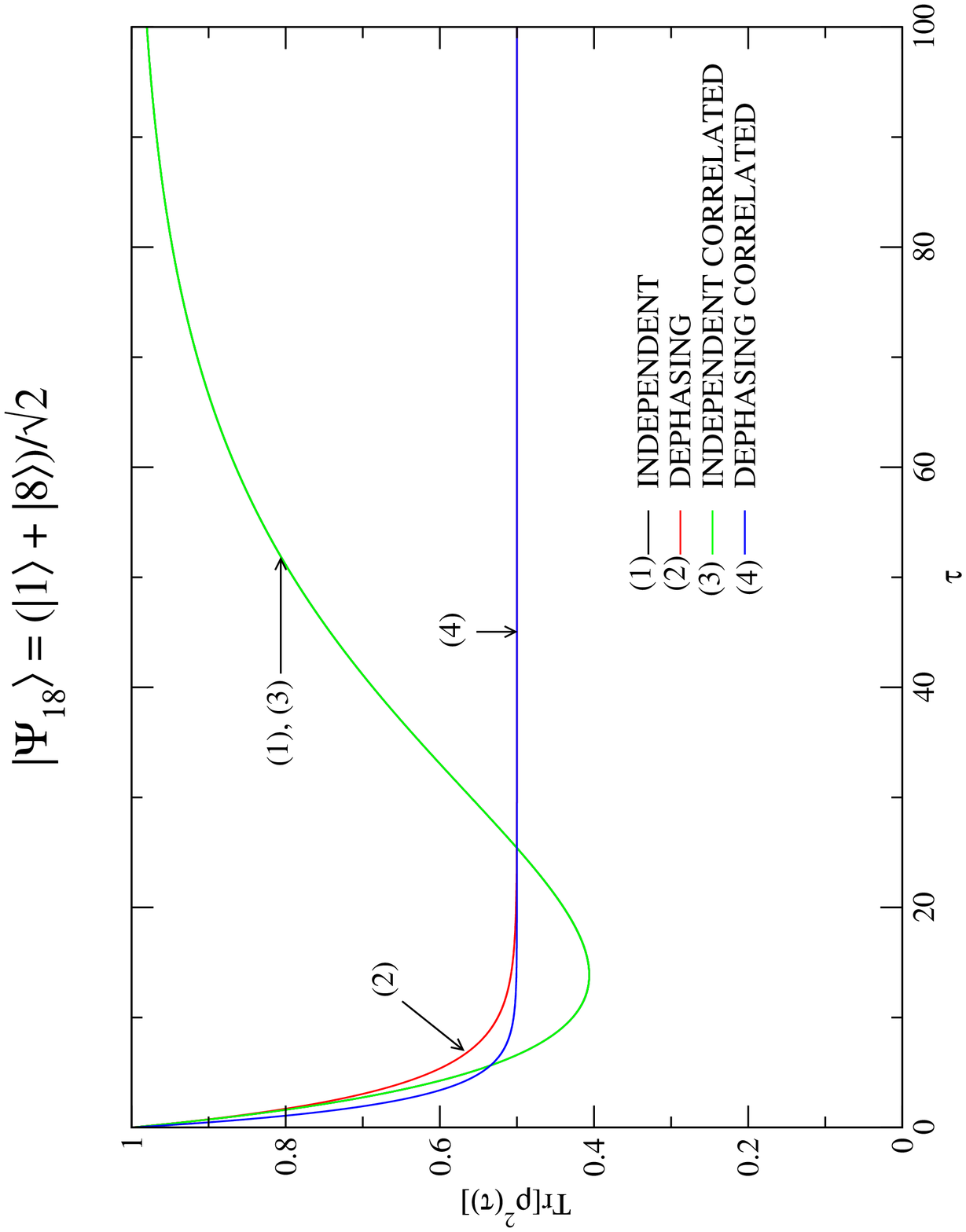}}
   \subfloat[]{
        \includegraphics[width=8cm,height=8cm]{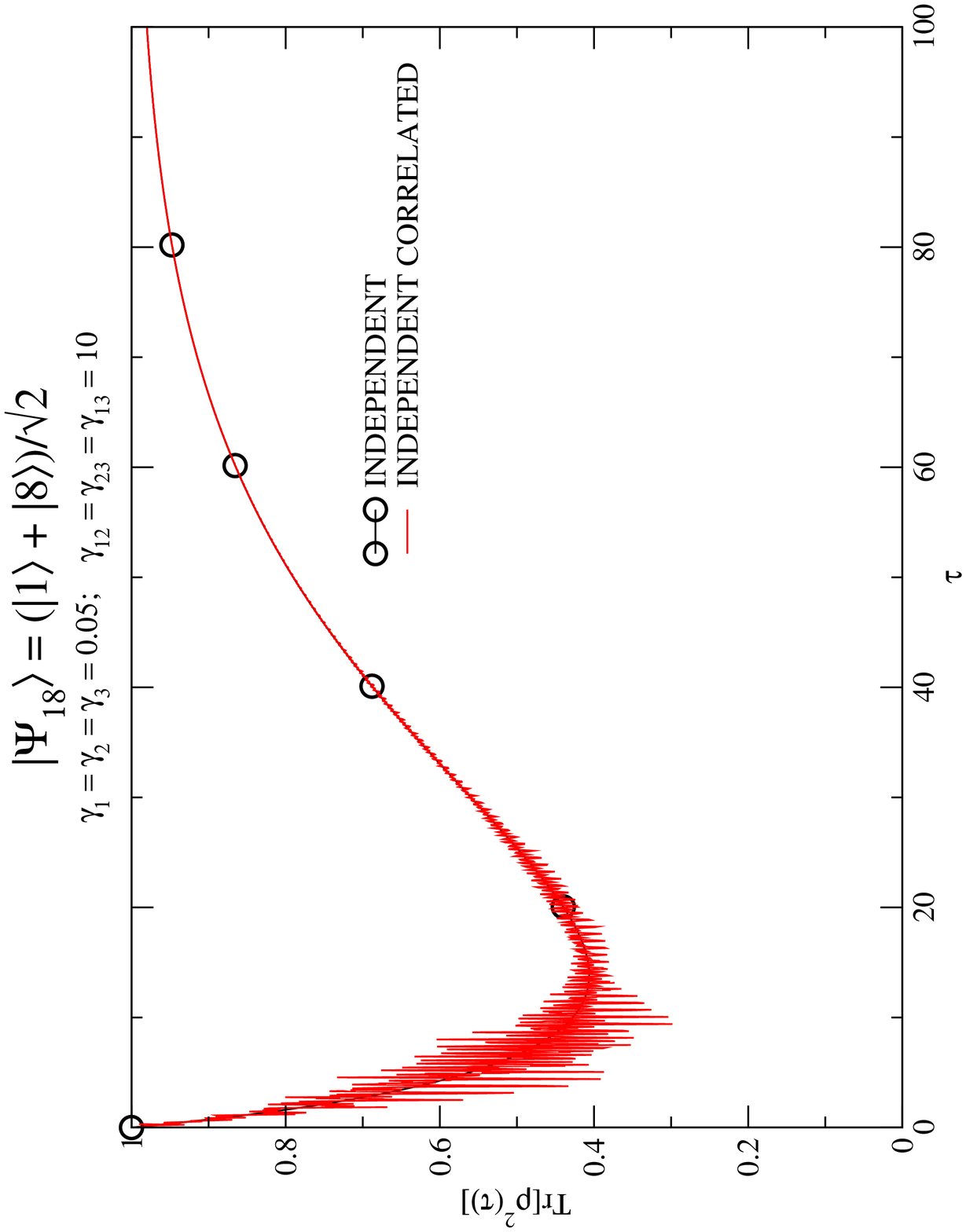}}
         \caption{\small{{\bf Independent Model:} (a) Purity behavior with 4-models environments. (b) Indepent model with big dissipation.}}
\end{figure}
\begin{figure}[h!]
   \centering
   \includegraphics[width=8cm,height=8cm]{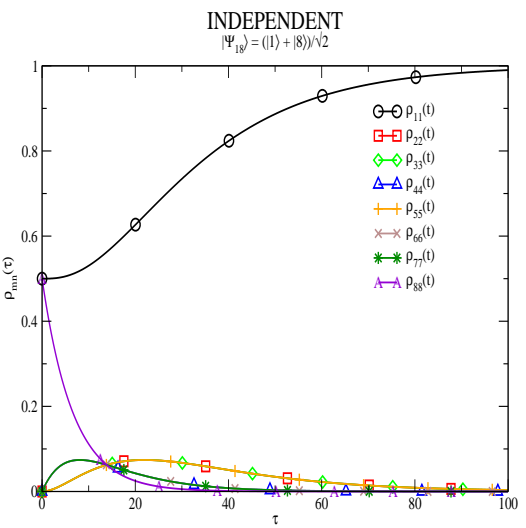}
    \caption{\small{Diagonal matrix elements behavior.}} 
\end{figure}
\noindent
Let us see now how the GME-concurrence and Purity behave for different entangled state and different environments.\\
{\bf Independent Model :} For the entangled states listed on table 1, Figures 3a, 3c, 4a, and 4c show GME-concurrence, and Figures 3b, 3d, 4b, and 4d show their associated Purity parameter behavior.  The system always finish on the pure ground states ($|1\rangle=|000\rangle$), the lowest bound of the GME-concurrence fall down (although this parameter can not tell us whether or not the entanglement has been  completely  destroyed) . Except for the entangled state $|\Psi_{18}\rangle$ (maximum entergy difference between their entangled qubits), there is not clear difference how this entanglement decay is developed. For example, the entangled states $|\Psi_{27}\rangle, |\Psi_{36}\rangle, |\Psi_{45}\rangle$ have the same GME-concurrence decay behavior, but these states have different energy-difference on their associated qubits.  The entangled states $|\alpha_{17}\rangle$ and $|\alpha_{46}\rangle$ the GME-concurrence decay is the same, although their energy-difference is quite big.\\ \\
\begin{figure}[ht!]
   \centering
   \subfloat[]{
        \includegraphics[width=8cm,height=7cm]{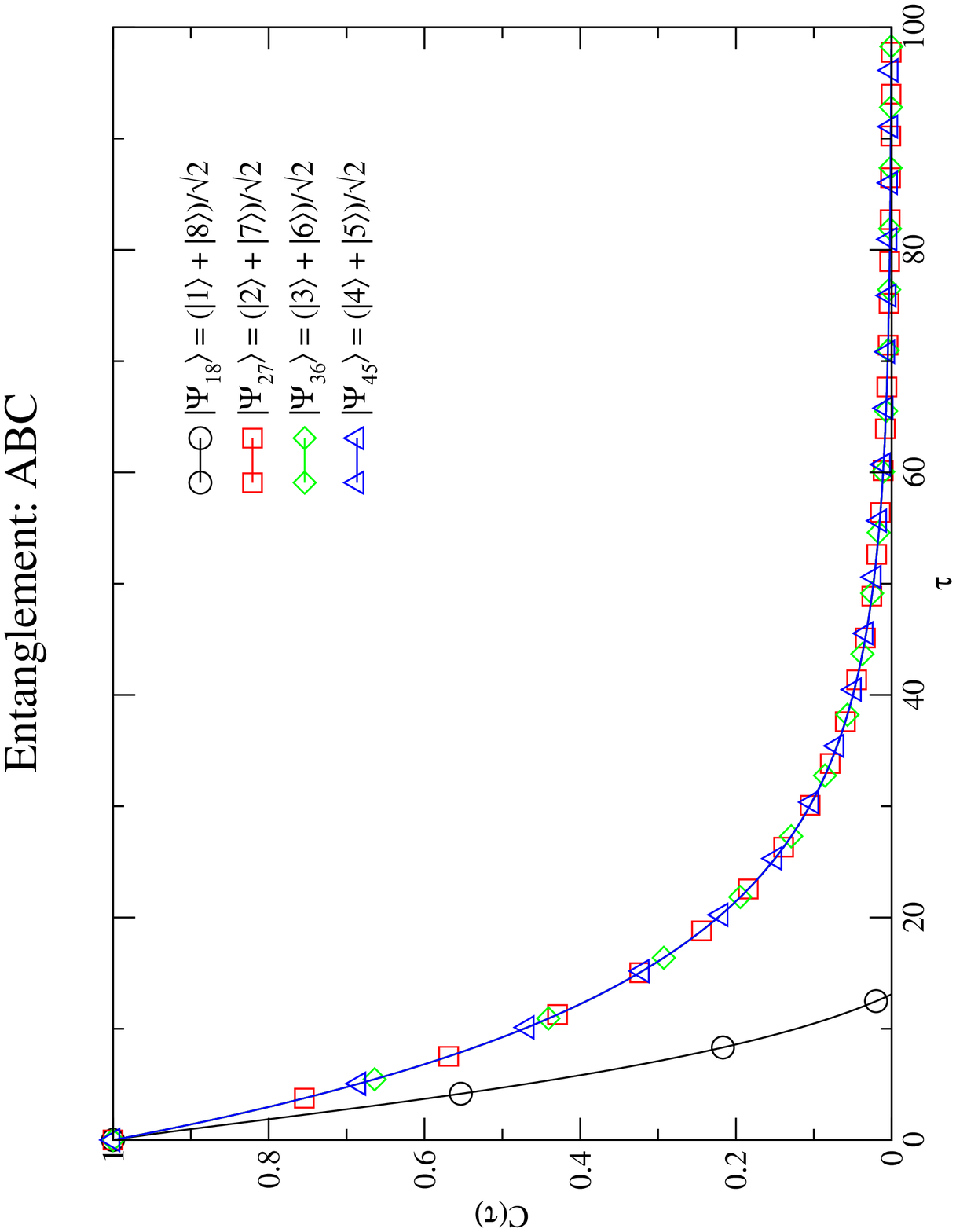}}
   \subfloat[]{
        \includegraphics[width=8cm,height=7cm]{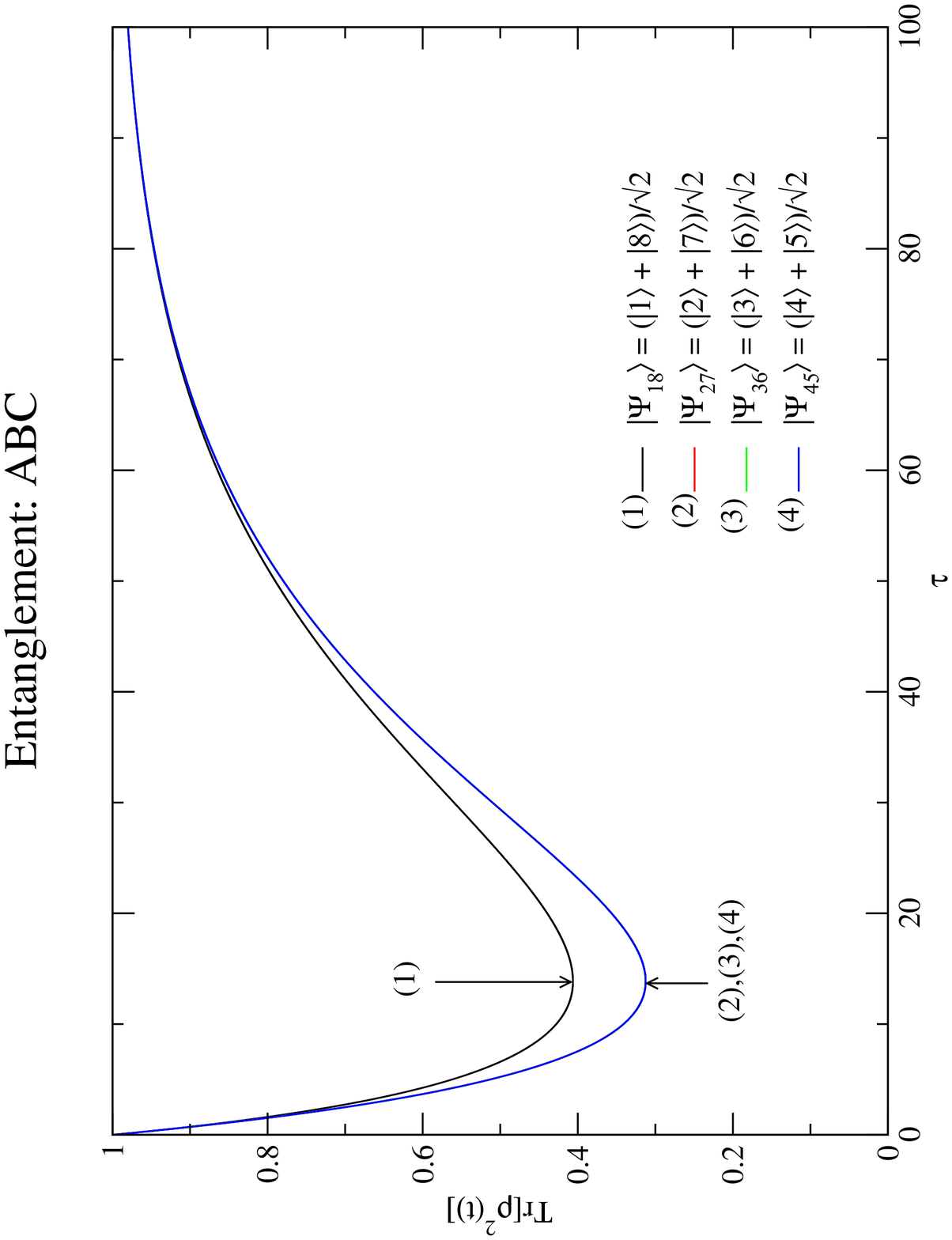}}\\[10pt]
  \caption{\small{Independent:(a) GME-concurrence $vs$ $\tau$. (b)  Purity $vs$ $\tau$.}}
\end{figure}
\newpage
\begin{figure}[ht!]
   \centering
   \subfloat[]{

        \includegraphics[width=8cm,height=8cm]{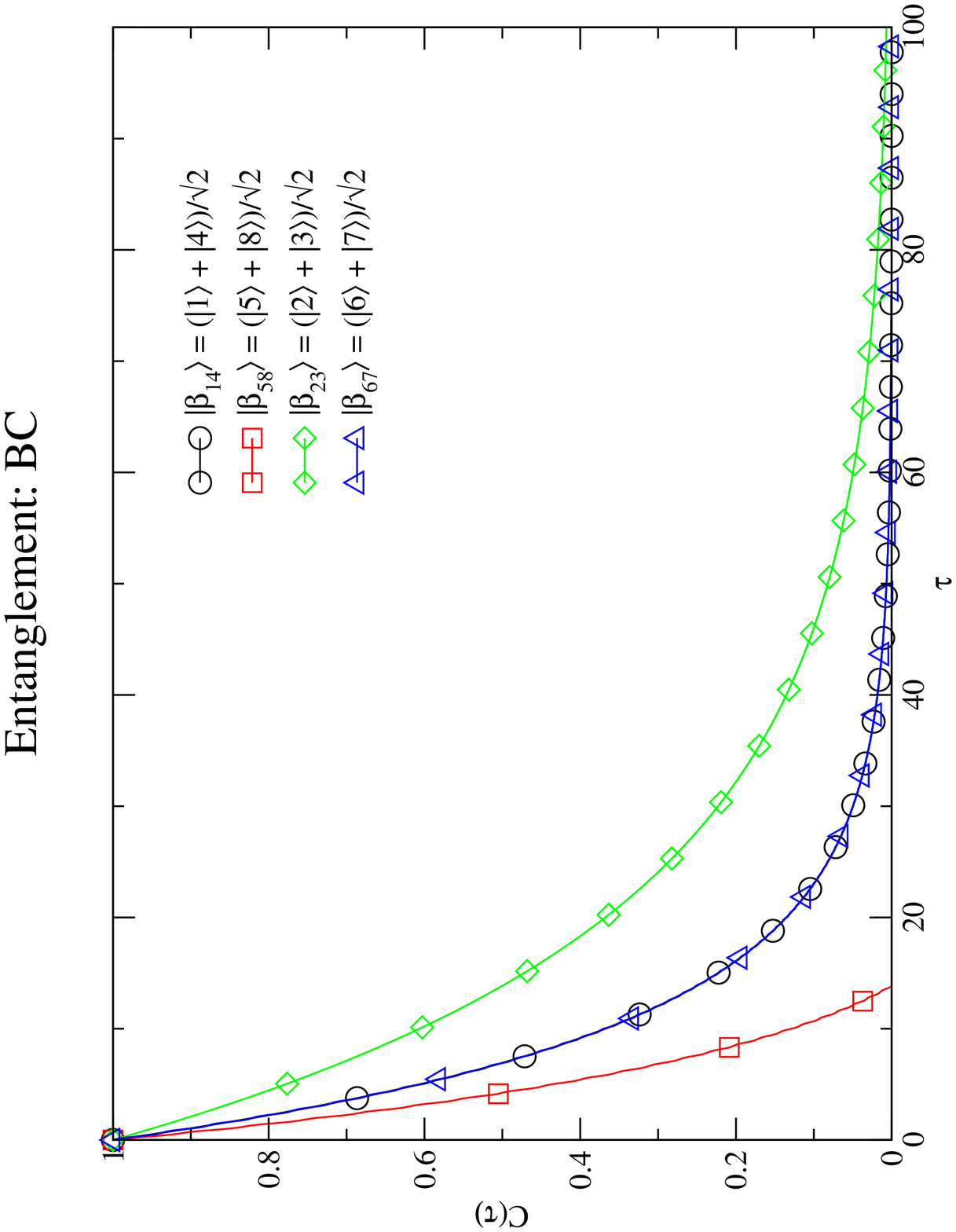}}
   \subfloat[]{
        \includegraphics[width=8cm,height=8cm]{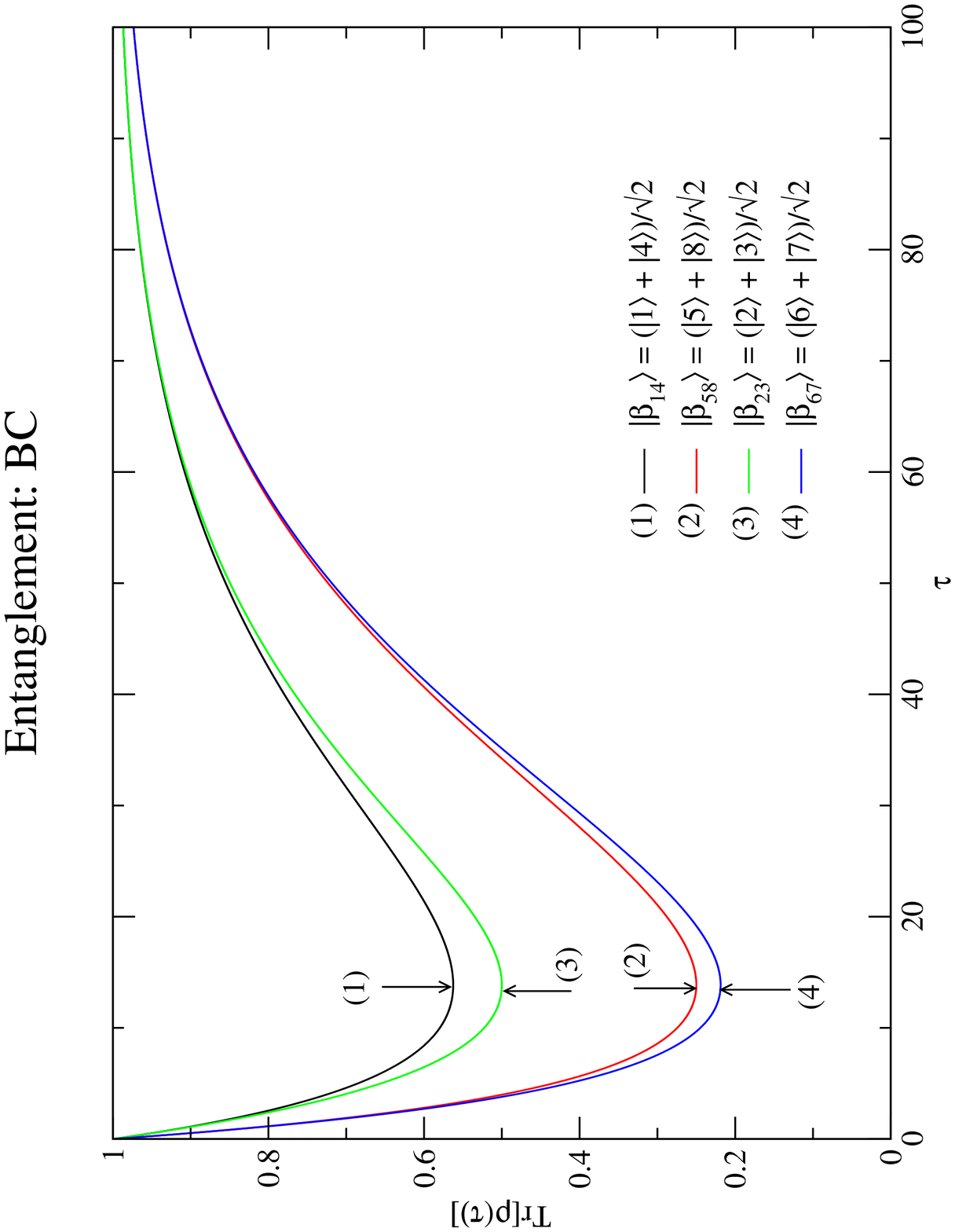}}\\[10pt]
  \caption{\small{Independent:(a) GME-concurrence $vs$ $\tau$. (b)  Purity $vs$ $\tau$.}} 
\end{figure}
%
\newpage
\noindent
{\bf Dephasing Model:}
In this case, the GME-concurrence and purity parameters can be expressed explicitly in a simple form as shown in the next table: 
\begin{table}[h!]
   \centering
   \caption{\small{}}
   \begin{tabular}{lll}
      Entanglement  &  $C_{GME:|\Upsilon_{ij}\rangle}(\tau)$    & $P_{|\Upsilon_{ij}\rangle}(\tau)$ \\\hline
   $(ABC)$   & $2|\rho_{ij}(0)\exp(-\Gamma \tau)|$,               & $\rho_{ii}^2(0) + \rho_{jj}^2(0) + 2\rho_{ij}^2(0)\exp(-\Gamma \tau)$ \\
  $(AB)$    & $2|\rho_{ij}(0)\exp[-(\Gamma_1 + \Gamma_2)\tau]|$, & $\rho_{ii}^2(0) + \rho_{jj}^2(0) + 2\rho_{ij}^2(0)\exp[-(\Gamma_1 + \Gamma_2)\tau]$ \\
 $(BC)$    & $2|\rho_{ij}(0)\exp[-(\Gamma_2 + \Gamma_3)\tau]|$, & $\rho_{ii}^2(0) + \rho_{jj}^2(0) + 2\rho_{ij}^2(0)\exp[-(\Gamma_2 + \Gamma_3)\tau]$ \\
  $(AC)$    & $2|\rho_{ij}(0)\exp[-(\Gamma_1 + \Gamma_3)\tau]|$, & $\rho_{ii}^2(0) + \rho_{jj}^2(0) + 2\rho_{ij}^2(0)\exp[-(\Gamma_1 + \Gamma_3)\tau]$
   \end{tabular}
\end{table}
\\
\noindent
where  $\Gamma = \Gamma_1 + \Gamma_2 + \Gamma_3$, and $|\Upsilon_{ij}\rangle = \lla{|\Psi_{ij}\rangle, |\alpha_{ij}\rangle, |\beta_{ij}\rangle, |\xi_{ij}}\rangle$, for the entangled cases  $ABC$, $AB$, $BC$ and $AC$ respectively. This expressions show that the decay behavior is the same for each family of entangled states, that is, entangled states in the same family have the same decay behavior.\\Ê\\
{\bf Independent correlated:}
From Figure 1,  we saw that correlations have not effect on the purity. In addition, Figure 5 shows entangled states  in different environments where we see that the behavior of the GME-concurrence is the same for the independent and independent correlated models. 
\begin{figure}[ht!]
   \centering
   \subfloat[]{
        \includegraphics[width=8cm,height=7cm]{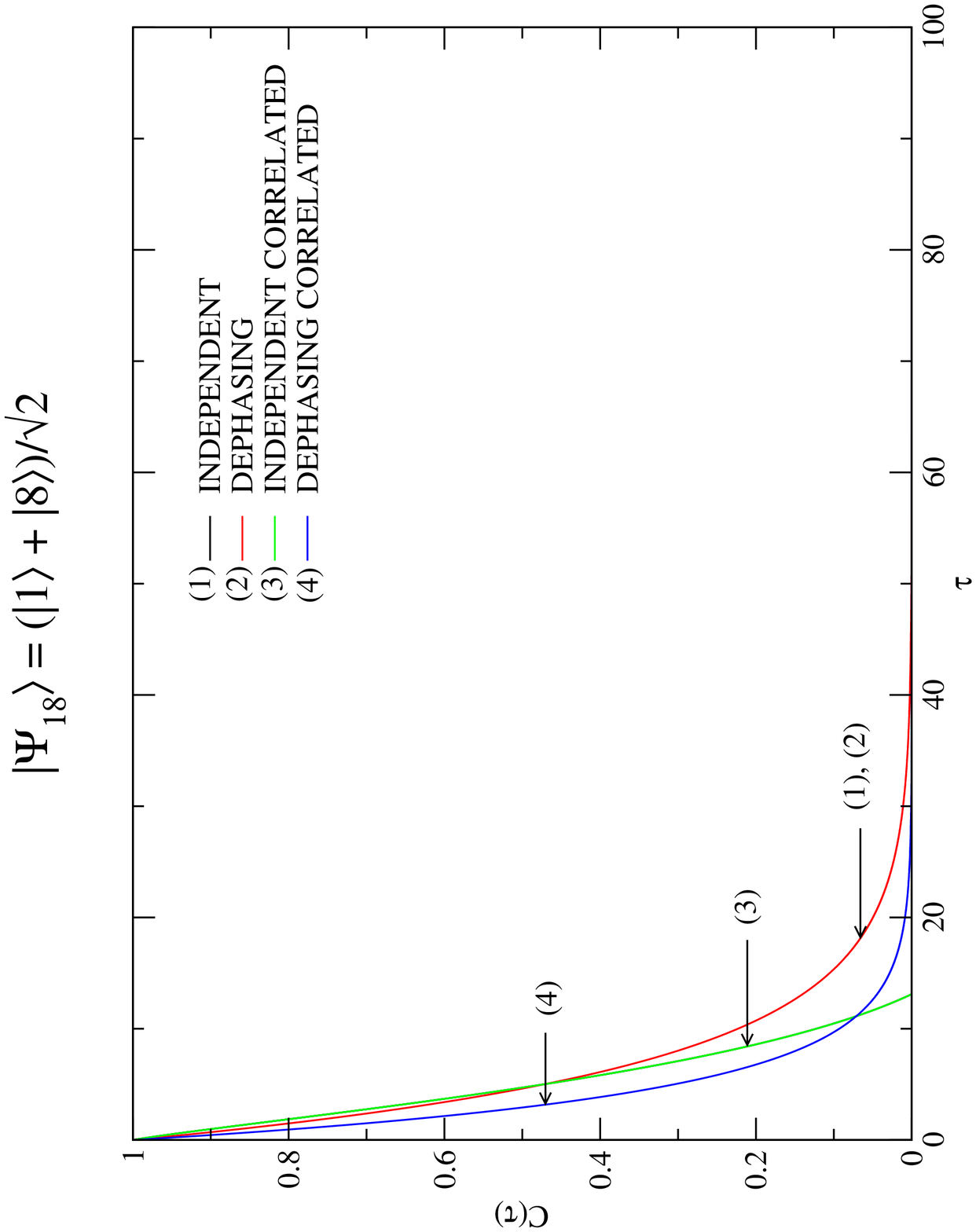}}
   \subfloat[]{
        \includegraphics[width=8cm,height=7cm]{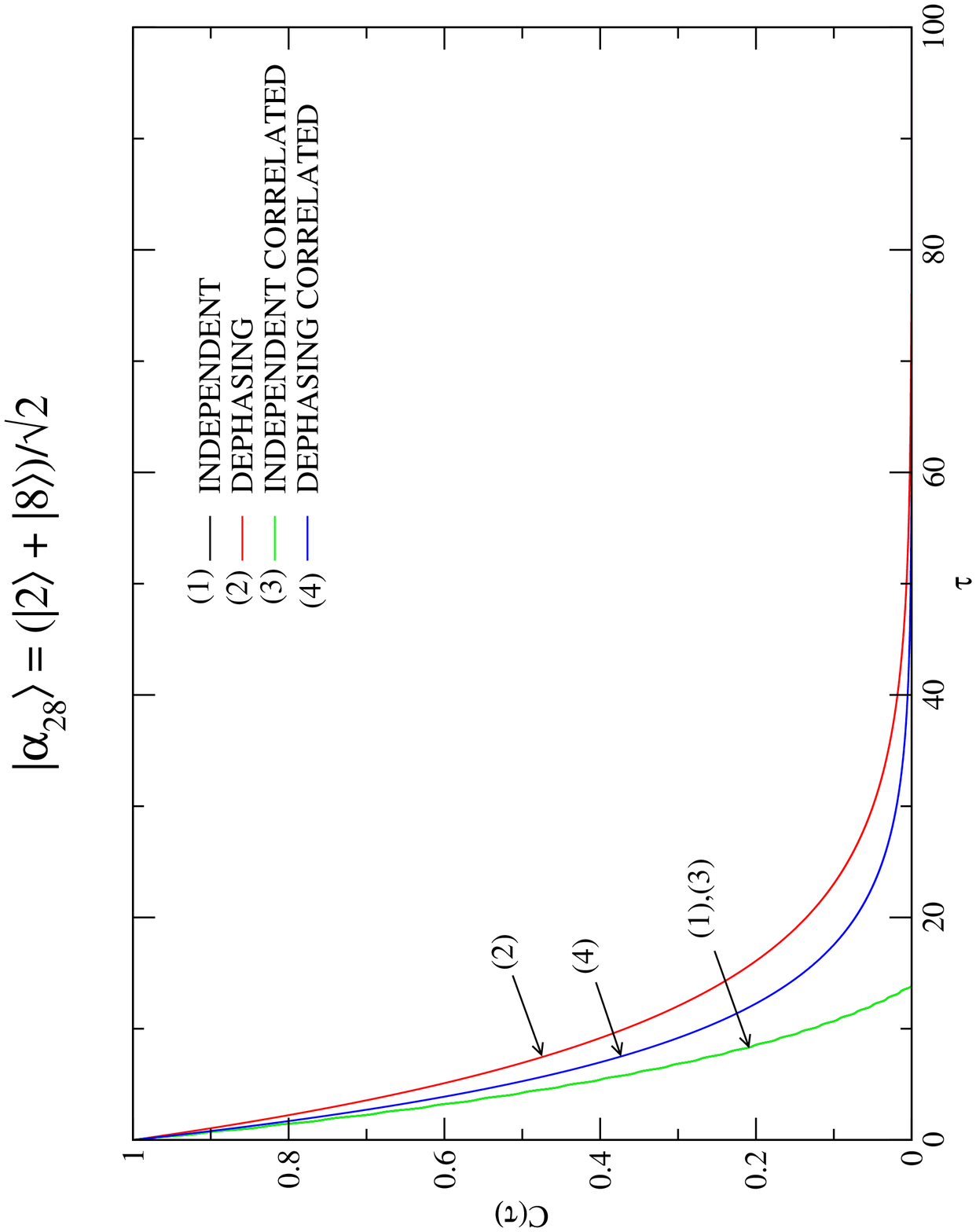}}\\[10pt]
  \caption{\small{GME-concurrence for entangled states with the four difference environments.}}
\end{figure}
%
     \newpage
   \section{Conclusions}
   We have studied the decay behavior of entangled states, formed by two basic states of three qubits registers, under four difference environments and using Lindblad type of equation to see whether or not this decay has a regular dependence with respect the energy-difference (difference of energy of the two basic states of three qubits which made up the entangled state) associated to the entangled state. We did not find this regular dependence, but rather a complicated situation which depends also on the type of environment.

\newpage\noindent
{\Large\bf Appendix}\\ \\
We consider a linear chain of three nuclear spin system. Then, our basis  is $\{|\xi_3\xi_2\xi_1\rangle\}_{\xi_i=0,1}$,  and the equations for the reduced matrix elements are obtained by making the bracket with these elements of the basis of the equation (\ref{LinU}).\\Ê\\
{\bf (a) Independent :}\\
\begin{equation}
\frac{\partial}{\partial t}\rho_{mn} + \sum_k\frac{\gamma_k}{2}\delta_{mn}(k)\rho_{mn} = \sum_k\gamma_k\tilde{\delta}_{mn}(k)e^{i\Delta\Omega_{mn}(k)t}\rho_{m+2^{N-k},n+2^{N-k}}
\end{equation}
where we have made the following definitions
\begin{eqnarray}
\delta_{mn}(k)         & = & \delta_{\alpha_k^{m},1}\delta_{\alpha_k^{m-2^{N-k}},0} + \delta_{\alpha_k^{n},1}\delta_{\alpha_k^{n-2^{N-k}},0}\\
\tilde{\delta}_{mn}(k) & = & \delta_{\alpha_k^{m},0}\delta_{\alpha_k^{n},0} \\
\Delta\Omega_{mn}(k)   & = & \Omega_{k,n+2^{N-k}} - \Omega_{k,m+2^{N-k}}.
\end{eqnarray}
{(b) Correlated independent :}\\
\begin{eqnarray}
\frac{\partial}{\partial t}\rho_{mn}(t) &=& \sum_{k,l}{N}\frac{\gamma_{kl}}{2}\left[2\delta_{mn}(k,l)e^{i\Delta\Omega_{mn}(k,l)t}\rho_{m+2^{N-k},n+2^{N-k}} \right.\nonumber\\
                                        & &        \hspace{1.25cm}-\delta_m(k,l)e^{i\Delta\Omega_{mn}'(k,l)t}\rho_{m-2^{N-l}+2^{N-k},n}      \nonumber\\
                                        & & \left. \hspace{1.25cm}-\delta_n(k,l)e^{i\Delta\Omega_{mn}''(k,l)t}\rho_{m,n-2^{N-l}+2^{N-k}}\right]    
\end{eqnarray}
where the following definitions have been made
\begin{align}
\delta_{mn}(k,l)& = \delta_{\alpha_{l}^{m},0}\delta_{\alpha_{k}^{n},0}         & \Delta\Omega_{mn}(k,l)  & = \Omega_{k,n+2^{N-k}} - \Omega_{l,m+2^{N-k}}\\
\delta_{m}(k,l) & = \delta_{\alpha_{l}^{m},1}\delta_{\alpha_{k}^{m-2^{N-l}},0} & \Delta\Omega_{mn}'(k,l) & = \Omega_{k,m-2^{N-l}+2^{N-k}} - \Omega_{l,m-2^{N-l}} \\
\delta_{n}(k,l) & = \delta_{\alpha_{l}^{n},1}\delta_{\alpha_{k}^{n-2^{N-l}},0} & \Delta\Omega_{mn}''(k,l)& = \Omega_{k,n-2^{N-l}+2^{N-k}} - \Omega_{l,n-2^{N-l}}
\end{align}
{\bf (c) Dephasing :}\\
\begin{equation}
\frac{\partial}{\partial t}\rho_{mn}(t) = \sum_k^N\Gamma_k\cor{(-1)^{\alpha_k^m + \alpha_k^n} - 1}\rho_{mn}(t)
\end{equation}
which has the following analytical solution
\begin{equation}
\rho_{mn}(t) = \rho_{mn}(0)\exp\lla{-\sum_k^N\Gamma_k\cor{1 - (-1)^{\alpha_k^m + \alpha_k^n}}t}.
\end{equation}
{\bf (d) Correlated depahsing :}\\
\begin{equation}                                
\frac{\partial}{\partial t}\rho_{mn}(t) = \sum_{k,l}^N\frac{\Gamma_{kl}}{4}\cor{2(-1)^{\alpha_l^m + \alpha_k^n} - (-1)^{\alpha_l^m + \alpha_k^m} - (-1)^{\alpha_l^n + \alpha_k^n} }\rho_{mn}(t)
\end{equation}
which has the explicit solution
\begin{equation}
\rho_{mn}(t) = \rho_{mn}(0)\exp\lla{-\sum_{k,l}^N\frac{\Gamma_{kl}}{4}\cor{(-1)^{\alpha_l^m + \alpha_k^m} + (-1)^{\alpha_l^n + \alpha_k^n} - 2(-1)^{\alpha_l^m + \alpha_k^n}}t}.
\end{equation}


\newpage
   \begin{thebibliography}{99}
      \bibitem{CaLeg}A.O. Caldeira and A.T. Legget, \ti{Path Integral Approach to Quantum Brownian Motion}, Physica A, {\bf 121}, No. 3, 587 (1983).
      \bibitem{HuPazZh}B.L. Hu, J.P. Paz, and Y. Zhang, \ti{Quantum Brawnian Motion in a General Environment: Exact Master Eqaution with Nonlocal Dissipation and Colored Noised}, Phys. Rev. D, {\bf 45}, No. 8, 2843 (1992).
      \bibitem{Venu}A. Venugopalan, \ti{Decoherence and Sch\"odinger-Cat State in a Stern-Gerlach-Type Experiment}, Phys. Rev. A, {\bf 56}, No. 5, 4307 (1997).
       \bibitem{Bre_OQS}Breuer, Heinz-Peter; F. Petruccione, \ti{The Theory of Open Quantum Systems}. Oxford University Press (2007).      
       \bibitem{Haro}S. Haroche and J.M. Raymond,  \ti{Exploring the Quantum}. Oxford University Press, London (2006).       
      \bibitem{ZuP}W.H. Zurek, Physics Today 44, 36 (1991).
      \bibitem{Zu2}W.H. Zurek, \ti{Decoherence, Einselection, and the Quantum Origens of the Classical}, Rev. Mod. Phys., {\bf 75}, No.3, 715 (2003).
      \bibitem{Zeh} H.D. Zeh, \ti{Toward Quantum Theory of Observation}, Foundation of Physics, {\bf 3}, No.1, 109 (1973).
      \bibitem{Zu1} W.H. Zurek, \ti{Decoherence and the Transition from Quantum to Classical}, Los Alamos Science, {\bf 27}, (2002).
      \bibitem{G_Lim} G. Lindblad, \ti{On the Generators of Quantum Dynamical Semigroups}, \ti{Commun. Math. Phys.} \tn{48}, pp. 119-130 (1976).
      \bibitem{All}P. Allard, M. Helgstrand, and T. H\"are, \ti{The Complete Homogeneous Master Equation for a Heteronuclear Two-Spin System in the Basis of Cartesian Product Operators}, Journal of Magnetic Resonace, {\bf 134}, No. 1, 7 (1998).      
      \bibitem{Suman}Das Sumanta  and G.S. Agarwal, \ti{Decoherence effects in interacting qubits under the influence of various environments}, J. Phys. B: At. Mol.  Opt.Phys., \tn{42} (2009).
      \bibitem{Berman_02} G.P. Berman, D.D. Doolen, D.I. Kamenev, G.V. L\'opez, V.I. Tsifrinovich, \ti{Perturbation Theory and Numerical Modeling of     Quantum Logic Operations with Large Number of Qubits},  Contemp. Math., \tn{305}, pp.13-41 (2002).        
      \bibitem{Lop_03} G.V. L\'opez, J. Quezada, G.P. Berman, D.D. Doolen, V.I. Tsifrinovich, \ti{Numerical simulation of a quantum controlled-not gate implemented on four-spin molecules at room temperature}, J. Opt. B: Quantum Semiclass Opt., \tn{5}, No.2, pp. 184-189 (2003).
      \bibitem{Lop_08} G.V.  L\'opez, T. Gorin, L. Lara, \ti{Simulation of Grover  quantum search algorithm in an Ising-nuclear-spin-chain quantum computer with first-and-second-nearest-neighbor couplings}, J. Phys. B: At. Mol. Opt. Phys., \tn{41}, No. 5, 055504 (2008).
      \bibitem{Lop_12} G.V. L\'opez, P. L\'opez, \ti{Study of Decoherence of Elementary Gates Implemented in a Chain of Few Nuclear Spins Quantum Computer Model }J. Mod. Phys., \tn{3}, pp. 85, (2012).      
      \bibitem{Lloyd_93} S. Lloyd, \ti{A potential Realizable Quantum Computer}, Science, \tn{261},  1569 (1993).
      \bibitem{Fano}U. Fano, \ti{Description of States in Quantum Mechanics by Density Matrix and Operator Techniques}, Rev. Mod. Phys., {\bf 29}, 74 (1957).  
      \bibitem{von} J. von Neumann, \ti{Wahrsheinlichkeitstheoretischer Aufbau der Quantenmechanik}, G\"ottinger Nachrichten, {\bf 1}, 245 (1927).    
      \bibitem{Alicki}R. Alicki, K. Lendi, \ti{Quantum Dynamical Semigroups and Applications}. Lecture Notes Phys., vol. 717. Springer, Berlin (2007).
       \bibitem{Davies}E.B. Davies, \ti{Quantum Theory of Open Systems}. Academic Press, San Diego (1976).
      \bibitem{Huber}M. Huber and F. Mintert, \ti{Detection of High-Dimensional Genuine Multipartite Entanglement of Mixed States}, Phys. Rev. Lett., \tn{104}, 210501 (2010).
      \bibitem{Seev}M. Seevinck and J. Uffink, \ti{Partial separability and entanglement criteria for multiqubit quantum states}, Phys. Rev. A, \tn{78}, 032101 (2008).
       \bibitem{Zhi}Zhi-Hao Ma \ti{et al.}, \ti{ Measure of genuine multipartite entanglement with computable lower bounds },  Phys. Rev. A, \tn{83}, 062325 (2011).
      \bibitem{NiCh}M. Nielsen and I. Chuang, \ti{Quantum Computation and Quantum Information}, Cambridge University Press, (2004).
      
   \end{thebibliography}
\end{document}